\documentclass[aps,pra,amsmath,amssymb,a4paper,reprint,longbibliography,showpacs,superscriptaddress]{revtex4-1}
\usepackage{graphicx}
\usepackage[breaklinks=true,colorlinks=true,citecolor=blue]{hyperref}
\usepackage{bm}

\newcommand{\mr}[1]{\mathrm{#1}}
\newcommand{\lp}{\left(}
\newcommand{\rp}{\right)}
\newcommand{\lam}{\lambda}
\newcommand{\lammax}{\lambda_\mr{max}}

\newcommand{\Tr}{\mathrm{Tr}}

\begin{document}

%%%%%%%%%%%%%%%%%%%%%%%%%%%%%%%%%%%%%%%%%%%%%%%%%
% Paper Information
%%%%%%%%%%%%%%%%%%%%%%%%%%%%%%%%%%%%%%%%%%%%%%%%%

\title{Capacity of entanglement and distribution of density matrix eigenvalues\\ in gapless systems}
\author{Yuya O. Nakagawa} 
\email{y-nakagawa@issp.u-tokyo.ac.jp}
\affiliation{Institute for Solid State Physics, University of Tokyo, 5-1-5 Kashiwanoha, Kashiwa, Chiba 277-8581, Japan}
\author{Shunsuke Furukawa}
\affiliation{Department of Physics, University of Tokyo, 7-3-1 Hongo, Bunkyo-ku, Tokyo 113-0033, Japan}
\date{\today}

% 03.65.Ud  Entanglement and quantum nonlocality
% 03.67.Mn  Entanglement measures, witnesses, and other characterizations
% 05.50.+q  Lattice theory and statistics (Ising, Potts, etc.)
% 05.70.Jk  Critical point phenomena
% 11.25.Hf Conformal field theory 
% 64.60.-i  General studies of phase transitions
% 71.10.Pm  Fermions in reduced dimensions (anyons, composite fermions, Luttinger liquid, etc.) 
% 71.10.Hf 	Non-Fermi-liquid ground states, electron phase diagrams and phase transitions in model systems 
% 71.27.+a 	Strongly correlated electron systems; heavy fermions
% 73.43.-f Quantum Hall effects
% 73.43.Cd Theory and modeling
% 75.10.Jm  Quantized spin models
% 75.10.Pq  Spin chain models
% 75.40.Mg  Numerical simulation studies

%%%%%%%%%%%%%%%%%%%%%%%%%%%%%%%%%%%%%%%%%%%%%%%%%
% Abstract
%%%%%%%%%%%%%%%%%%%%%%%%%%%%%%%%%%%%%%%%%%%%%%%%%

\begin{abstract}
We propose that the properties of the capacity of entanglement (COE) in gapless systems 
can efficiently be investigated through the use of the distribution of eigenvalues of the reduced density matrix (RDM). 
The COE is defined as the fictitious heat capacity calculated from the entanglement spectrum. 
Its dependence on the fictitious temperature can reflect the low-temperature behavior of the physical heat capacity, 
and thus provide a useful probe of gapless bulk or edge excitations of the system. 
Assuming a power-law scaling of the COE with an exponent $\alpha$ at low fictitious temperatures, 
we derive an analytical formula for the distribution function of the RDM eigenvalues. 
We numerically test the effectiveness of the formula in relativistic free scalar boson in two spatial dimensions,  
and find that the distribution function can detect the expected $\alpha=3$ scaling of the COE 
much more efficiently than the raw data of the COE. 
We also calculate the distribution function in the ground state of the half-filled Landau level with short-range interactions, 
and find a better agreement with the $\alpha=2/3$ formula than with the $\alpha=1$ one, 
which indicates a non-Fermi-liquid nature of the system. 
\end{abstract}
\maketitle

%%%%%%%%%%%%%%%%%%%%%%%%%%%%%%%%%%%%%%%%%%%%%%%%%
\section{Introduction} \label{Section_Intro}
%%%%%%%%%%%%%%%%%%%%%%%%%%%%%%%%%%%%%%%%%%%%%%%%%

% [ Entanglement ]--------------------
Quantum entanglement, which represents nonlocal correlations that cannot be described by classical mechanics, 
has played a central role in quantum information science, and recently become an indispensable tool in the studies of quantum many-body systems. 
One can extract various properties of a system by calculating entanglement measures 
in the many-body (mostly, ground-state) wave function $|\Psi\rangle$ \cite{Laflorencie2016,Amico2008}. 
The most celebrated measure among them is the entanglement entropy (EE). 
By partitioning the system into a subregion $A$ and its complement $\bar{A}$, 
the EE is defined as the von Neumann entropy $S_A:=-\Tr~ \rho_A \ln \rho_A$ 
of the reduced density matrix (RDM) $\rho_A: = \Tr_{\bar{A}} |\Psi\rangle \langle\Psi |$. 
When the ground state $|\Psi\rangle$ contains only short-range correlations, the EE scales with the boundary size of $A$ (boundary law) \cite{Srednicki1993,Eisert2010}. 
Deviation from a boundary law signals the presence of certain nontrivial correlations, 
and can furthermore reveal universal numbers characterizing the system. 
In one-dimensional (1D) quantum critical systems, for example, the EE for an interval of length $x$ shows a logarithmic scaling $S_A=\frac{c}{3}\log \frac{x}{a}$, 
where $c$ and $a$ are the central charge and the (non-universal) short-distance cutoff of 
underlying conformal field theory (CFT) \cite{Holzhey1994, Vidal2003, Calabrese2004,Calabrese2009review}. 
In noninteracting fermions and Fermi liquids, the EE can detect a Fermi surface 
through a multiplicative logarithmic correction to a boundary law \cite{Wolf2006,GioevKlich2006,Swingle2010,Swingle2012,Ding2012,McMinis2013}. 
Interestingly, the EE can also detect a hidden Fermi surface of emergent particles (such as spinons and composite fermions) 
in a similar manner \cite{ZhangGrover2011,Grover2013,Swingle2013,Shao2015,Mishmash2016,LaiYang2016}, 
providing a guiding principle for constructing a holographic dual of a strongly interacting metal \cite{Ogawa2012,Huijse2012}.  
In topologically ordered systems \cite{KitaevPreskill2006, LevinWen2006, Hamma2005, Hamma2005PhysLettA} and in some 2D critical systems \cite{Metlitski2009,HsuFradkin2009,Stephan2009}, 
the EE obeys a boundary law, but there appears a subleading universal constant that reflects underlying topological or critical properties. 
While the EE was initially featured on the theoretical side, 
state-of-the-art techniques in ultracold atomic systems can now measure it experimentally \cite{MeasurementEE2015, Science2016, Hannes2016}, 
fostering further growing interest among both theorists and experimentalists. 

% [ Entanglement spectrum ]--------------------
Since the EE can be calculated from the eigenvalues of the RDM, 
the latter can in principle contain more information of the system than the former. 
This idea has led to the notion of entanglement spectrum (ES) \cite{LiHaldane2008}. 
By rewriting the RDM in the thermal form $\rho_A=e^{-H_E}$, 
where $H_E$ is referred to as the entanglement Hamiltonian, 
the ES is defined as the full eigenvalue spectrum of $H_E$. 
Although the ES is calculated from the ground state, 
a number of studies have demonstrated that the ES resembles the physical energy spectrum of the system. 
In gapped topological phases, in particular, 
the ES has been found to exhibit the same low-energy features as the physical edge-mode spectrum 
\cite{KitaevPreskill2006, LiHaldane2008, Thomale2010, Hong2010, TurnerZhangVishwanath2010, Fidkowski2010, Sterdyniak2012, Dubail2012RSES}. 
Several physical ``proofs'' have been given for this remarkable correspondence 
\cite{QiKatsuraLudwig2012,Chandran2011,Dubail2012,SwingleSenthil2012,Lundgren2013,CanoHughes2015} 
while some exceptions to it have also been discussed \cite{Chandran2014, Ho2015}. 

% [ Entanglement spectrum in gapless systems ]--------------------
The correspondence between the ES and the physical spectrum has also been found in some gapless systems. 
In 1D critical systems, beautiful numerical evidences have been presented 
for the correspondence between the ES and the energy spectrum of a boundary CFT \cite{Lauchli2013}. 
In systems with spontaneous continuous symmetry breaking, the ES has been found to exhibit a tower structure 
in a way analogous to the physical spectrum \cite{MetlitskiGrover2011,Alba2013, Kolley2013}. 
In gapless phases of spin ladders, however, the ES between the chains has been found to exhibit a flat or fractional dispersion relation 
as opposed to a linear energy dispersion of a single chain \cite{ChenFradkin2013,Lundgren2013}.

% [ Capacity of entanglement ]--------------------
To gain further insights into the properties of the ES, 
it is useful to look into the ``thermodynamics'' of the entanglement Hamiltonian $H_E$. 
The capacity of entanglement (COE) has been introduced for such a purpose \cite{Hong2010,Schliemann2011,Nakaguchi2016}. 
The COE $C_E(T_E)$ is defined as the fictitious heat capacity of $H_E$, where $T_E$ is the fictitious temperature 
(see Sec.~\ref{Section_Derivation} for a precise definition of the COE). 
The correspondence between the ES and the physical spectrum 
can then be revealed by the correspondence between the COE and the physical heat capacity. 
In 1D critical systems, the CFT prediction $\Tr~\rho_A^n\sim (x/a)^{\frac{c}{6}(n-1/n)}$ \cite{Holzhey1994,Calabrese2004,Calabrese2009review} 
leads to a linear scaling $C_E\sim T_E$ \cite{Nakaguchi2016,note_COE_CFT}, 
which coincides with the low-temperature behavior of the physical heat capacity \cite{Affleck1986}. 
Free fermions and Fermi liquids with a Fermi surface can be described as a collection of CFTs \cite{Swingle2010,Swingle2012,Ding2012}, 
and thus the COE of these systems is also expected to show a linear scaling $C_E\sim T_E$ at low $T_E$ as the physical heat capacity does. 
In more general gapless systems, the correspondence between the COE and the physical heat capacity is unclear, 
and some counterexamples to the correspondence are known \cite{Nakaguchi2016}. 
However, one can still use the COE to probe unusual low-energy properties of the system. 
For example, based on the above consideration, 
a non-Fermi-liquid behavior can be signaled by the violation of the linear scaling of the COE 
(see also Ref.~\cite{Swingle2013} for a related discussion). 
This indicates an advantage of the COE over the EE 
as the latter does not seem to distinguish Fermi and non-Fermi liquids in a qualitative manner \cite{ZhangGrover2011,Grover2013,Shao2015,Mishmash2016}. 
Furthermore, the COE has an advantage over the physical heat capacity in that the former requires only the ground-state wave function 
and can be applied to a trial wave function. 

 % [ This paper ]--------------------
In this paper, we investigate the behaviors of the COE and the distribution of the ES (more precisely, the distribution of the RDM eigenvalues) in some gapless systems. 
We find that a nontrivial low-$T_E$ behavior of the COE can efficiently be detected through the use of the distribution of the ES. 
Specifically, by assuming a power-law behavior $C_E\sim T_E^\alpha$ at low $T_E$, 
we derive an analytic formula for the cumulative distribution function $n(\lam)$ of the RDM eigenvalues [see Eq.~\eqref{Formula} below]. 
This is based on a generalization of the work by Calabrese and Lefevre for 1D critical systems \cite{Calabrese2008}. 
We numerically test the effectiveness of the formula in relativistic free scalar boson in two spatial dimensions,  
and find that $n(\lam)$ can detect the expected $\alpha=3$ scaling of the COE 
much more efficiently than the raw data of the COE. 
This advantage of $n(\lam)$ results from a sensitive dependence of the analytic formula \eqref{Formula} on $\alpha$.  
As a more nontrivial application, we then study the half-filled Landau level with short-range interactions. 
For this system, Halperin, Lee, and Read (HLR) \cite{HLR1993} formulated a theory of a Fermi sea of composite fermions 
(see also Refs.~\cite{Barkeshli2015,Murthy2016,Son_PRX2015,Wang2016,Wang2016_2,Geraedts2016,Levin2017,Wang2017} 
for recent interesting theoretical developments on this system). 
Gauge fluctuations in the HLR theory were shown to make a singular contribution to a heat capacity, 
which scales as $T^{2/3}$ if the bare interaction between fermions is short-range \cite{HLR1993,KimLee1996}. 
We have calculated $n(\lam)$ of this system by using the ground state obtained by exact diagonalization, 
and find a better agreement with the $\alpha=2/3$ formula than with the $\alpha=1$ one, 
which indicates a non-Fermi-liquid nature.  
While our data obtained for maximally $N=14$ particles do not allow a precise determination of $\alpha$, 
a relatively good agreement with the $\alpha=2/3$ formula suggests an intriguing possibility 
that the correspondence between the ES and the physical spectrum still holds in a strongly interacting metallic state.

% [ Organization of the paper ]--------------------
The rest of the paper is organized as follows.
In Sec.~\ref{Section_Derivation}, we derive the analytical formula for the distribution of the RDM eigenvalues by assuming a power-law behavior of the COE. 
In Sec.~\ref{Section_Numerics},  we present numerical results in free scalar boson and the half-filled Landau level. 
In Sec.~\ref{Section_Conclusion},  we conclude the paper and discuss implications of our study.

%% ------ Section 2 ---------------------------- 
%%%%%%%%%%%%%%%%%%%%%%%%%%%%%%%%%%%%%%%%%%%%%%%%%
\section{Capacity of entanglement and distribution of density matrix eigenvalues\label{Section_Derivation}} 
%%%%%%%%%%%%%%%%%%%%%%%%%%%%%%%%%%%%%%%%%%%%%%%%%

% [ Section introduction ]--------------------
In this section, we first describe the definitions of the COE and the distribution of the RDM eigenvalues. 
We then derive an analytic formula for the distribution of the RDM eigenvalues 
by assuming a power-law behavior of the COE, $C_E\sim T_E^\alpha$, at low $T_E$. 

%************************************************
\subsection{Definitions}
%************************************************

% [ Capacity of entanglement ]--------------------
Let us first clarify the definitions of the COE and the distribution of the RDM eigenvalues. 
Using the RDM $\rho_A=e^{-H_E}$ on a subregion $A$, we introduce the entanglement partition function as
\begin{equation}\label{DefZE}
 Z_E (T_E) := \Tr~e^{-H_E/T_E} = \Tr~\rho_A^{1/T_E}. 
\end{equation}
The COE is then defined as \cite{Hong2010,Schliemann2011,Nakaguchi2016}
\begin{equation}\label{DefCOE}
 C_E (T_E) :=  T_E \frac{\partial^2}{\partial T_E^2} [T_E \ln Z_E(T_E)]. 
\end{equation}
In the above expressions, we dropped the dependence on $A$ as it is not considered throughout our analysis. 
We are instead interested in the dependence on the fictitious temperature $T_E$. 
As the entanglement Hamiltonian $H_E$ is dimensionless, so is $T_E$. 
We note that studying the dependence of the COE on $T_E$ is equivalent to studying the R\'enyi EE 
\begin{equation*}
S_n := \frac{-1}{n-1} \ln R_n,~~R_n:=\Tr~\rho_A^n =Z_E (1/n)
\end{equation*}
as a function of the R\'enyi parameter $n$. 

% [ Distribution of RDM eigenvalues ]--------------------
Next we introduce the distribution of the RDM eigenvalues. 
We denote the eigenvalues of the RDM $\rho_A$ by $\{ \lambda_i \}$.
Since $\rho_A$ is positive semidefinite and has unit trace, these eigenvalues satisfy $0 \le \lam_i \le 1$ and $\sum_i \lam_i = 1$. 
The distribution function $P(\lam)$ and the cumulative distribution function $n(\lam)$ of $\{ \lambda_i \}$ are defined as \cite{Calabrese2008}
\begin{align}
 P(\lam) := \sum_i \delta (\lam -\lam_i), \:\: n(\lam) := \int_\lam^{\lammax}  P(\lam) d\lam,
\end{align}
where $\lammax$ is the largest eigenvalue.
Here, $n(\lam)$ counts the number of eigenvalues in the range $[ \lam, \lammax]$.
If $\{\lambda_i\}$ is sorted in descending order ($\lambda_1=\lammax\ge \lambda_2\ge \lambda_3\ge \dots$), 
$n(\lam)$ can also be viewed as the inverse function of $\lambda_i$. 

%************************************************
\subsection{Derivation of an analytic formula}\label{Subsection_Derivation}
%************************************************

% [ Subsection introduction ]--------------------
By assuming $C_E\sim T_E^{\alpha} $ with $\alpha > 0$ at low $T_E$, 
we now derive an analytic formula for the cumulative distribution function $n(\lambda)$. 
Our derivation is based on a generalization of the argument by Calabrese and Lefevre for 1D critical systems \cite{Calabrese2008} 
where $C_E\sim T_E$ at low $T_E$. 

% [ Entanglement partition function ]--------------------
Since $Z_E(T_E)$ is related with the COE via Eq.~\eqref{DefCOE}, 
our assumption on the COE immediately leads to 
\begin{align*}
  \ln Z_E(T_E) = b T_E^\alpha + b' + b''/T_E, 
\end{align*}
where $b$, $b'$ and $b''$ are constants. 
One can determine these constants by using some properties of $Z_E(T_E)$. 
In the limit $T_E \to 0$, the definition of $Z_E(T_E)$ in Eq.\ \eqref{DefZE} yields $Z_E(T_E) \to \lammax^{1/T_E} $, 
which indicates $b'=0$ and $b'' = \ln \lammax$. 
By further using $Z_E(T_E=1) = \Tr~\rho_A  =1$, we find $b = - b'' = - \ln \lammax$.
We thus obtain a simple form
\begin{align} \label{FormZET}
 \ln Z_E(T_E)  = b (T_E^\alpha-1/T_E),~b=-\ln \lammax.
\end{align}

% [ Calculate of n(\lambda) ]--------------------
Following Ref.\ \cite{Calabrese2008}, we introduce the function 
\begin{equation}\label{f_z}
 f(z) :=  \frac{1}{\pi} \sum_{n=1}^\infty R_n z^{-n} = \frac{1}{\pi} \int d\lam' \frac{\lam' P(\lam')}{z-\lam'},
\end{equation}
with which the distribution function $P(\lam)$ can be obtained as  $\lam P(\lam) = \lim_{\epsilon \to +0 } \mathrm{Im} f(\lam- i\epsilon)$.
By using Eq.~\eqref{FormZET}, we can calculate $f(z)$ as
\begin{align} 
 f(z) 
&= \frac{1}{\pi} \sum_{n=1}^\infty e^{ b(1/n^\alpha-n)} z^{-n} \nonumber \\
&= \frac{1}{\pi} \sum_{n=1}^\infty \lp \frac{\lammax}{z}\rp^n \sum_{k=0}^\infty \frac{1}{k!} \lp \frac{b}{n^\alpha} \rp^k   \nonumber \\
&= \frac{1}{\pi} \sum_{k=0}^\infty \frac{b^k}{k!} \mr{Li}_{\alpha k}(\lammax/z), \label{PolyLogExpression}
\end{align}
where $\mr{Li}_m(y) = \sum_{n=1}^\infty y^n / n^m$ is the polylogarithm function.
The function $\mr{Li}_m(y) \: (m > 0)$ has a branch cut on the real axis for $y \geq 1$ with the discontinuity  
\begin{equation}
\lim_{\epsilon \to +0}  \mr{Li}_m(y + i \epsilon) - \mr{Li}_m(y) =
\begin{cases}
 i \pi \frac{(\ln y)^{m-1}}{ \Gamma(m)} & (m>0) ; \\
 i\pi \delta(1-y) & (m=0),
\end{cases}
\end{equation} 
as noted in Ref.\ \cite{Calabrese2008}.
Therefore, by taking the limit $z \to \lam - i0$ in Eq.~\eqref{PolyLogExpression}, we obtain
\begin{align*}
 \lam P(\lam)  =  \lammax \delta( \lam - \lammax)  + \Theta(\lammax - \lam)  \sum_{k=1}^\infty
 \frac{b^k \Lambda^{\alpha k -1}}{k! \: \Gamma(\alpha k)}, 
\end{align*}
where $\Lambda = \ln(\lammax/\lam)$ and $\Theta(x)$ is the Heaviside step function.
By integrating $P(\lam)$, we arrive at the formula
\begin{align} \label{Formula}
 n (\lam) = 1 + \sum_{k=1}^\infty
 \frac{[ b (\ln(\lammax/\lam))^\alpha ]^k}{k! ~\Gamma(\alpha k+1)},
\end{align}
which plays a central role in this paper. 
Although this formula is expressed as an infinite series, 
it can be evaluated numerically for given $\alpha>0$ as it converges rapidly. 
When $\alpha=1$, the infinite series can be rewritten as the modified Bessel function, 
resulting in the formula of Calabrese and Lefevre \cite{Calabrese2008}. 
We note that when $\alpha$ is a rational number, $n_\alpha(\lam)$ can be written as
a sum of the generalized hypergeometric functions (see Appendix~\ref{Appendix_hypergeo}).

%% -------------- Section 3 -------------------
%%%%%%%%%%%%%%%%%%%%%%%%%%%%%%%%%%%%%%%%%%%%%%%%%
\section{Numerical results \label{Section_Numerics}}
%%%%%%%%%%%%%%%%%%%%%%%%%%%%%%%%%%%%%%%%%%%%%%%%%

In this section, we present numerical results on the COE $C_E(T_E)$ and the cumulative distribution function $n(\lam)$ of the RDM eigenvalues in some gapless systems. 
We first test the effectiveness of the formula \eqref{Formula} in relativistic free scalar boson in two spatial dimensions. 
Then, as a more nontrivial application, we present an exact diagonalization result in the half-filled Landau level with short-range interactions. 
By comparing the numerical data with the formula \eqref{Formula}, we find a signature of a non-Fermi-liquid nature of this system. 

%************************************************
\subsection{Relativistic free scalar boson in two spatial dimensions \label{subsec_boson} }
%************************************************

%############################
\begin{figure}
 \includegraphics[width = 8.5cm]{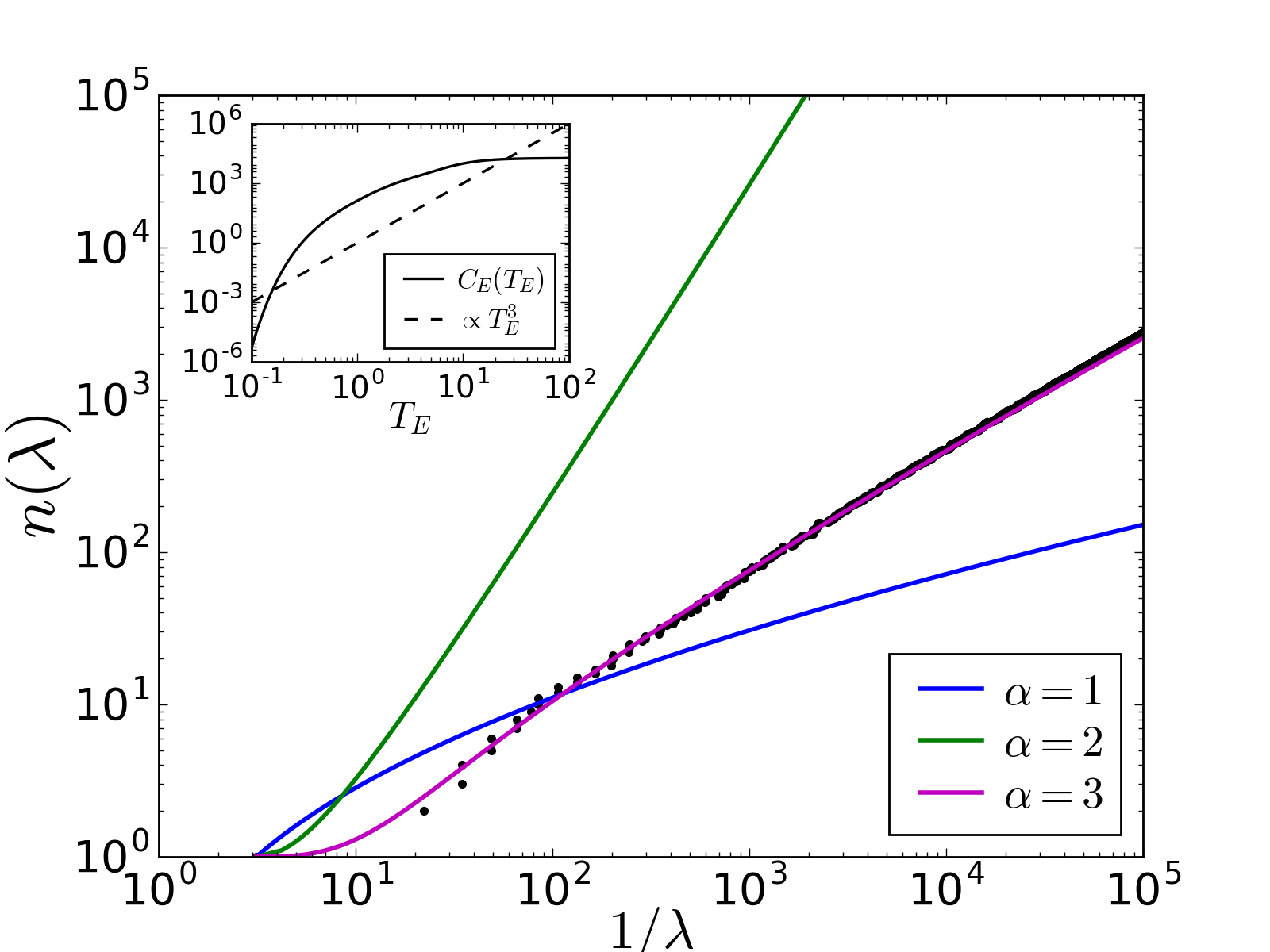}
 \caption{(color online) Cumulative distribution function $n(\lam)$ of the RDM eigenvalues in relativistic free scalar boson in two spatial dimensions.
 In numerical calculations, we discretized the field theory and took as the subregion $A$ a circle with a radius of $10$ discretized units.  
 Lines indicate the analytic formula \eqref{Formula} for three different values of the exponent $\alpha$. 
 We find a good agreement of the numerical data with the $\alpha=1$ formula. 
 The inset shows the COE $C_E(T_E)$ (solid line) in comparison with the expected low-$T_E$ scaling $C_E\propto T_E^3$ \cite{Nakaguchi2016} (dashed line). }
 \label{COE_boson2d}
\end{figure}
%############################

% [ System ]--------------------
Building on the analytic expression of the R\'enyi EE obtained by Klebanov {\it et al.} \cite{Klebanov2012}, 
Nakaguchi and Nishioka \cite{Nakaguchi2016} have calculated the COE of relativistic massless free scalar boson . 
In two spatial dimensions, in particular, the COE has been shown to scale as $C_E\sim T_E^3$ at low $T_E$. 
Interestingly, this is different from the low-$T$ behavior of the physical heat capacity $C\sim T^2$, 
providing a counterexample to the correspondence between the two quantities. 

% [ Numerical technique ]--------------------
An advantage of this system for testing the formula \eqref{Formula} is the availability of an efficient numerical technique for computing the RDM eigenvalues. 
Following Ref.~\cite{Nakaguchi2015}, we discretize the field theory of scaler boson $\phi$ with the action 
\begin{equation}
 S = \int d^2xdt \left[ (\partial_t \phi)^2 - (\nabla \phi)^2 \right],
\end{equation} 
and calculate the RDM eigenvalues by taking as a subregion $A$ a circle centering at the origin. 
Further technical details of the numerical calculation are described in Appendix~\ref{Appendix_boson}. 

% [ Numerical result ]--------------------
Figure~\ref{COE_boson2d} presents the cumulative distribution function $n(\lam)$ and the COE $C_E(T_E)$ (inset) calculated numerically.
It is clear that the data of $n(\lam)$ agree well with the analytic formula with $\alpha=3$ as expected. 
In contrast, the data of $C_E(T_E)$ plotted in logarithmic scales show a significant variation of slope; 
estimation of $\alpha$ through the fitting with the form $C_E\sim T_E^\alpha$ 
would crucially depend on the range of $T_E$ used for the fitting. 
These results indicate an advantage of  $n(\lam)$ over the COE $C_E(T_E)$ in determining the exponent $\alpha$. 
This advantage results from a sensitive dependence of the formula \eqref{Formula} on $\alpha$.

%************************************************
\subsection{Half-filled Landau level}
%************************************************

%############################
\begin{figure*}[bth]
 \includegraphics[width=8cm]{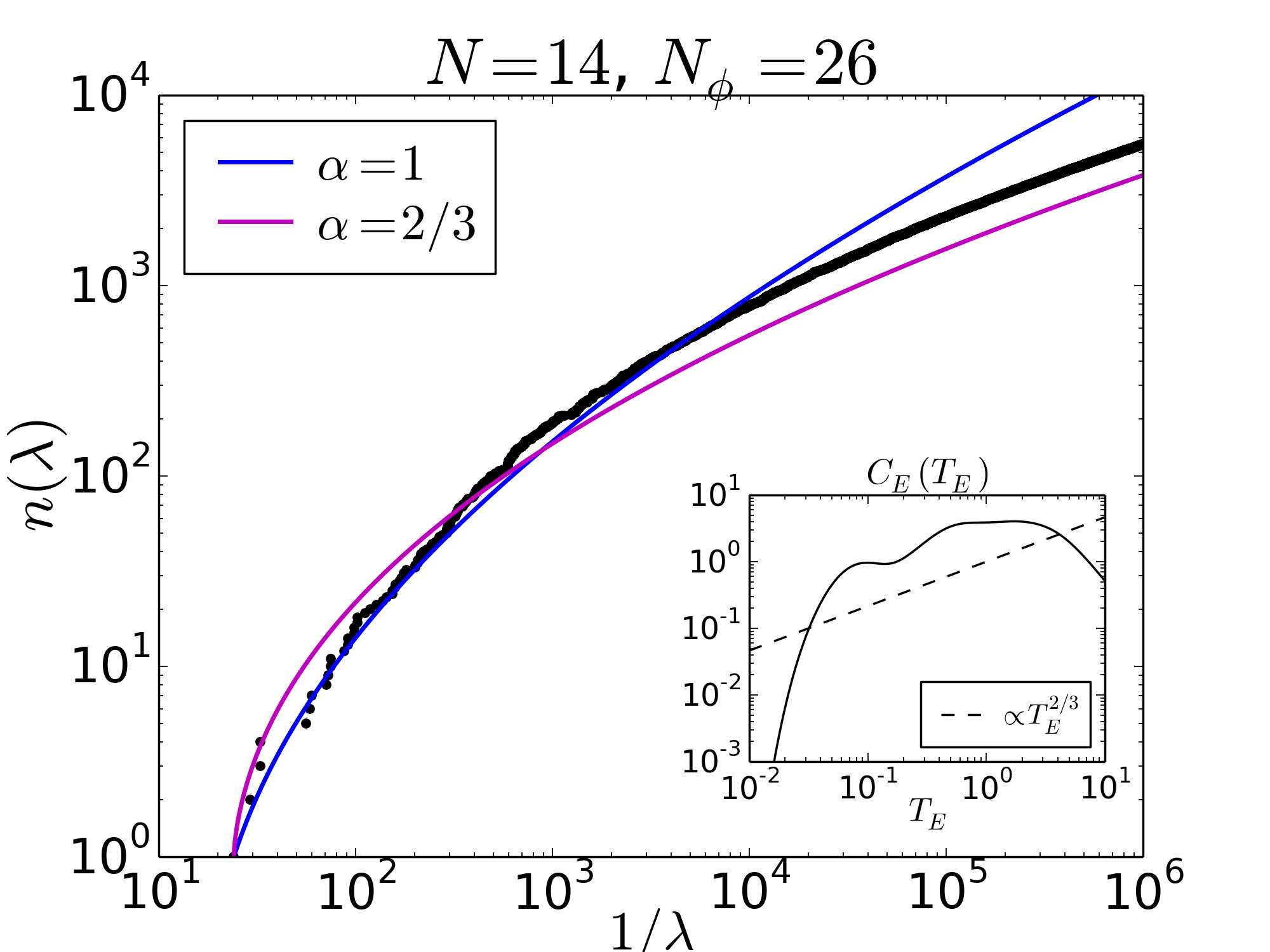}
 \includegraphics[width=8cm]{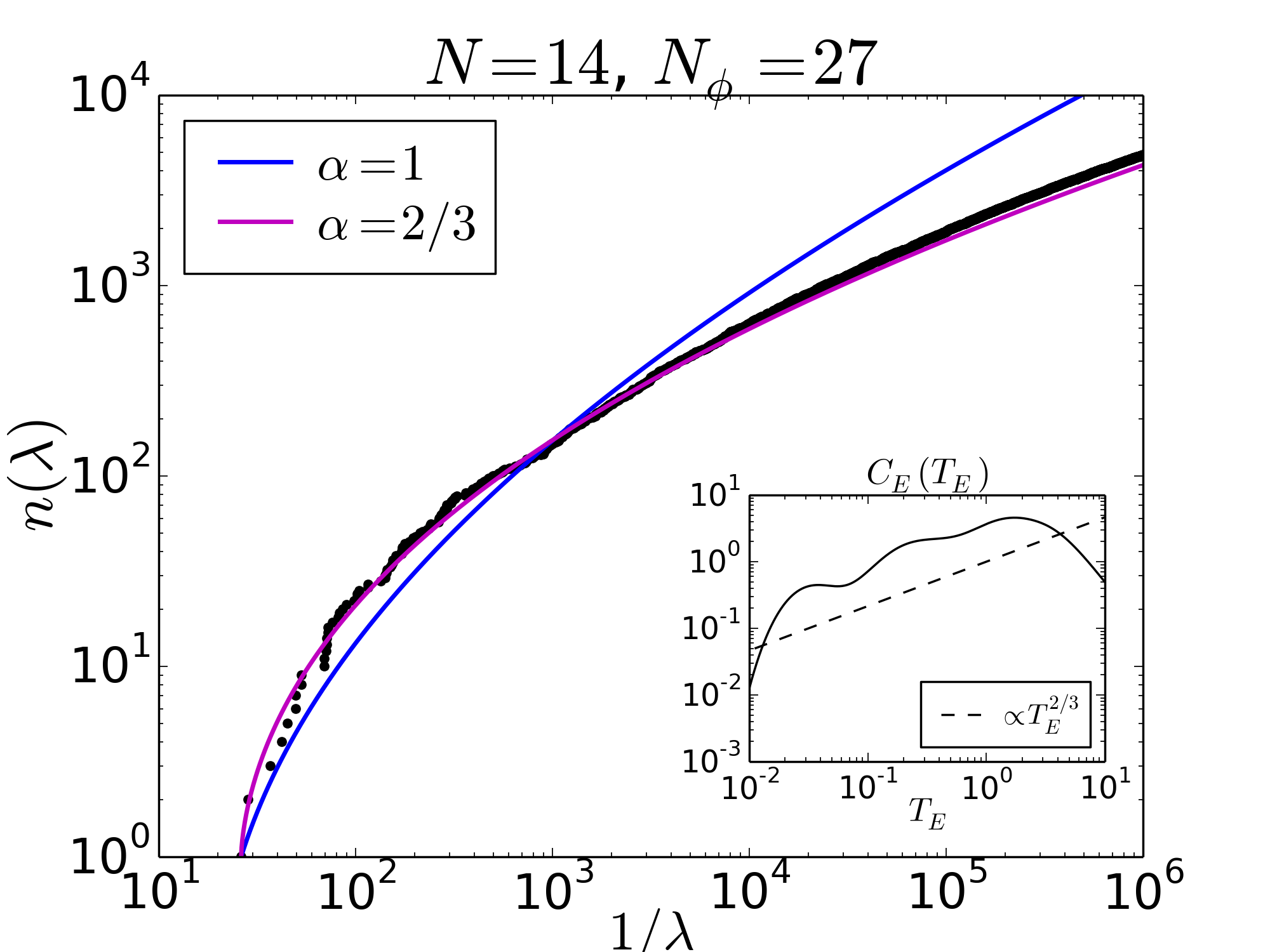}
 \caption{(color online) Cumulative distribution function $n(\lam)$ of the RDM eigenvalues for the ground state of the half-filled Landau level  (dots), 
 in comparison with the analytic formula \eqref{Formula} for $\alpha=2/3$ and $\alpha=1$ (lines). 
 The left and right panels correspond to the Fermi sea case ($N_\phi= 2N-2$) and the particle-hole symmetric case ($N_\phi= 2N-1$), respectively, where $N$ is set to $14$. 
 In plotting the analytical formula, the numerically obtained value of $\lambda_\mathrm{max}$ is used. 
 The insets show the COE $C_E(T_E)$ (solid line) in comparison with the power-law behavior $C_E\propto T_E^{2/3}$ (dashed line).
}
 \label{Comparison_fig}
\end{figure*}
%############################

% [ System ]--------------------
As a more nontrivial application of our formula~\eqref{Formula},
we consider a quantum Hall system at the filling factor $\nu=1/2$ (half-filled Landau level). 
For this system, HLR formulated an effective field theory 
in which composite fermions forms a Fermi sea and interact via a Chern-Simons gauge field \cite{HLR1993}. 
Gauge fluctuations in this theory were shown to make a singular contribution to a heat capacity \cite{HLR1993,KimLee1996}. 
In particular, when the bare interaction between fermions is short-range, 
the heat capacity was predicted to scale as $C\sim T^{2/3}$, indicating a non-Fermi-liquid behavior. 
However, this prediction has not been verified numerically 
as a large number of low-lying eigenenergies are required to obtain a low-temperature behavior of the heat capacity. 
Recently, there have been very active studies on the role of particle-hole symmetry in this system. 
The HLR theory does not satisfy this symmetry, and an alternative description in terms of Dirac composite fermions 
consistent with this symmetry has been developed \cite{Son_PRX2015,Wang2016,Wang2016_2,Geraedts2016,Levin2017}. 
In this description, Dirac composite fermions have a Fermi surface and interact via a gauge field without a Chern-Simons term. 
While the heat capacity has not been calculated in the Dirac scenario, the gauge field coupled with a Fermi surface is still expected to make a significant contribution. 

% [ Previous studies ]--------------------
Concerning entanglement properties, the $n=2$ R\'enyi EE has recently been calculated for trial wave functions of the half-filled Landau level, 
and a multiplicative logarithmic correction to the boundary law, which indicates a hidden Fermi surface, has been verified \cite{Shao2015, Mishmash2016}. 
However, the R\'enyi EE for fixed $n$ could not reveal a non-Fermi-liquid nature of the system. 
This motivates us to investigate the COE and the distribution of RDM eigenvalues in this system. 

% [ Numerical method ]--------------------
We performed exact diagonalization calculation for interacting $N$ spinless fermions 
in the lowest Landau level on a spherical geometry \cite{Haldane1983}. 
In this geometry, a magnetic monopole of charge $N_\phi$ in units of the flux quantum $h/e$ is placed at the center. 
We assume a repulsive short-range interaction $V(\bm{r})=-\nabla^2\delta(\bm{r})$ between fermions;  
this interaction is equivalent to the Haldane's pseudopotential for $\nu=1/3$ Laughlin state \cite{Haldane1983,Trugman1985}
while we here focus on the filling $\nu=1/2$. 
A Fermi sea of composite fermions corresponds to a set of $(N,N_\phi)$ satisfying $N_\phi = 2N-2$ \cite{Rezayi1994} 
whereas the particle-hole symmetric state (with a possible Dirac nature) is consistent with those satisfying $N_\phi = 2N-1$ \cite{Son_PRX2015}. 
We investigate both types of states in numerical calculations; in the thermodynamic limit, they both correspond to the filling factor $\nu=1/2$. 
The ground state is $(2L+1)$-fold degenerate if it has a total angular momentum of magnitude $L>0$; 
in such a case, we took the ground state in the $L_z=0$ sector for the computation of the RDM, 
where $L_z$ is the $z$-component of the total angular momentum.  
From such a ground state, we calculated the eigenvalues of the RDM associated with the real-space cut into two hemispheres \cite{Sterdyniak2012,Dubail2012RSES}. 

\begin{figure*}
 \includegraphics[width = 8.5cm]{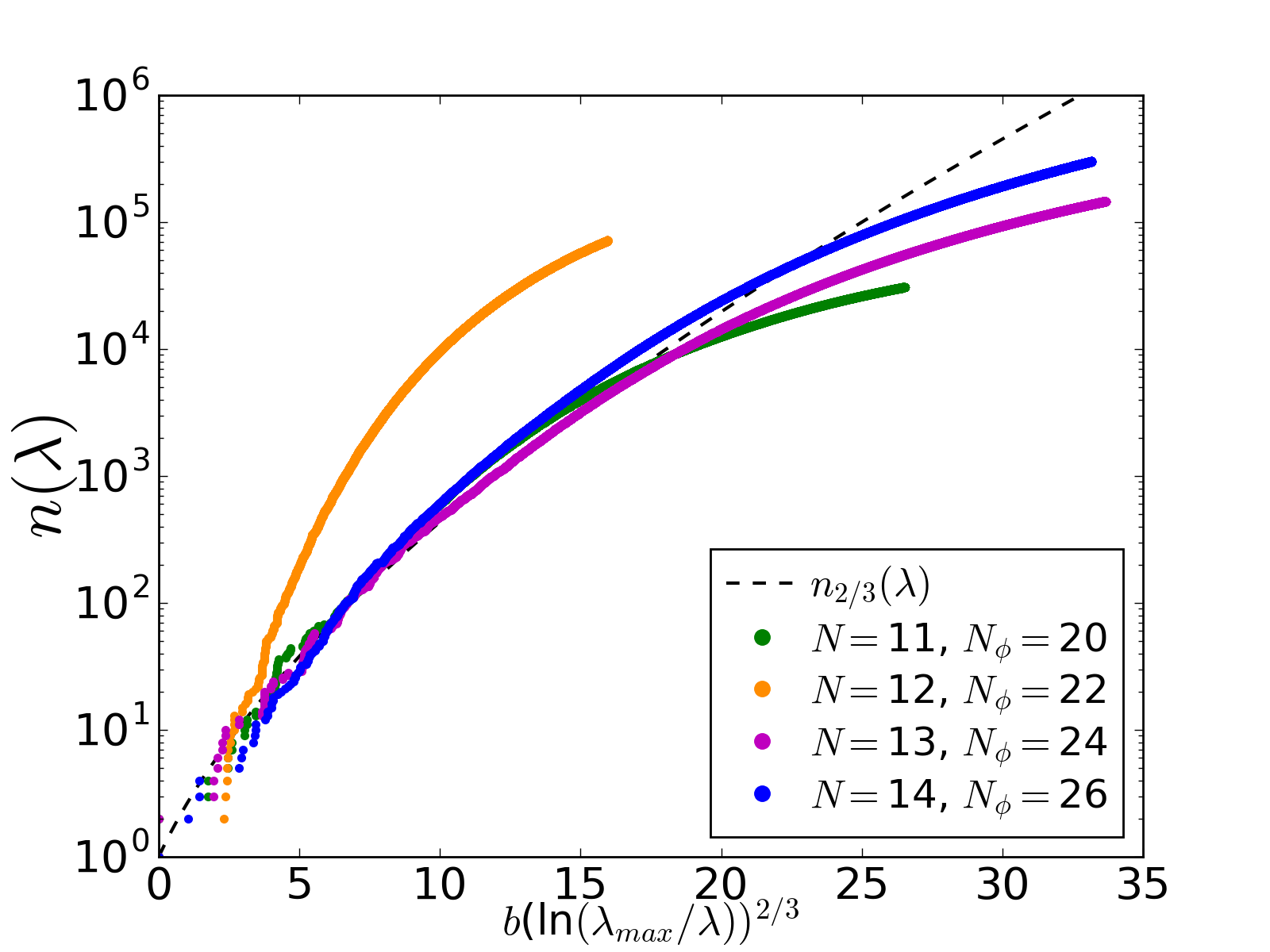}
 \includegraphics[width = 8.5cm]{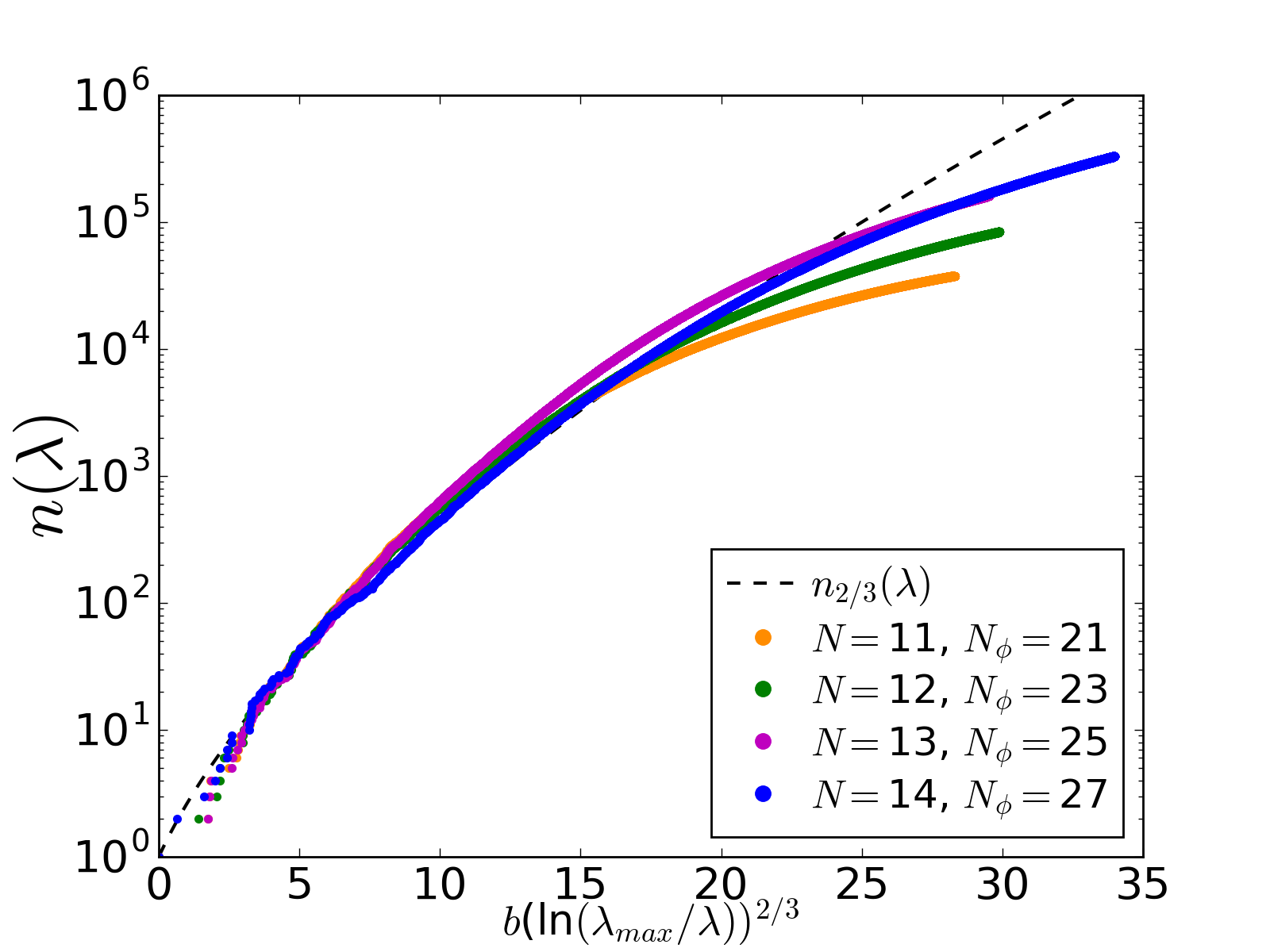}
 \caption{(color online) Cumulative distribution function $n(\lam)$ for the ground state of the half-filled Landau level for different $N$. 
 Numerical results (colored dots) are compared with the analytic formula \eqref{Formula} with $\alpha=2/3$ (dashed line). 
 The left and right panels correspond to the Fermi sea case ($N_\phi= 2N-2$) and the particle-hole symmetric case ($N_\phi= 2N-1$), respectively.}
 \label{COE_HFLL}
\end{figure*}

\begin{figure*}
 \includegraphics[width = 8.5cm]{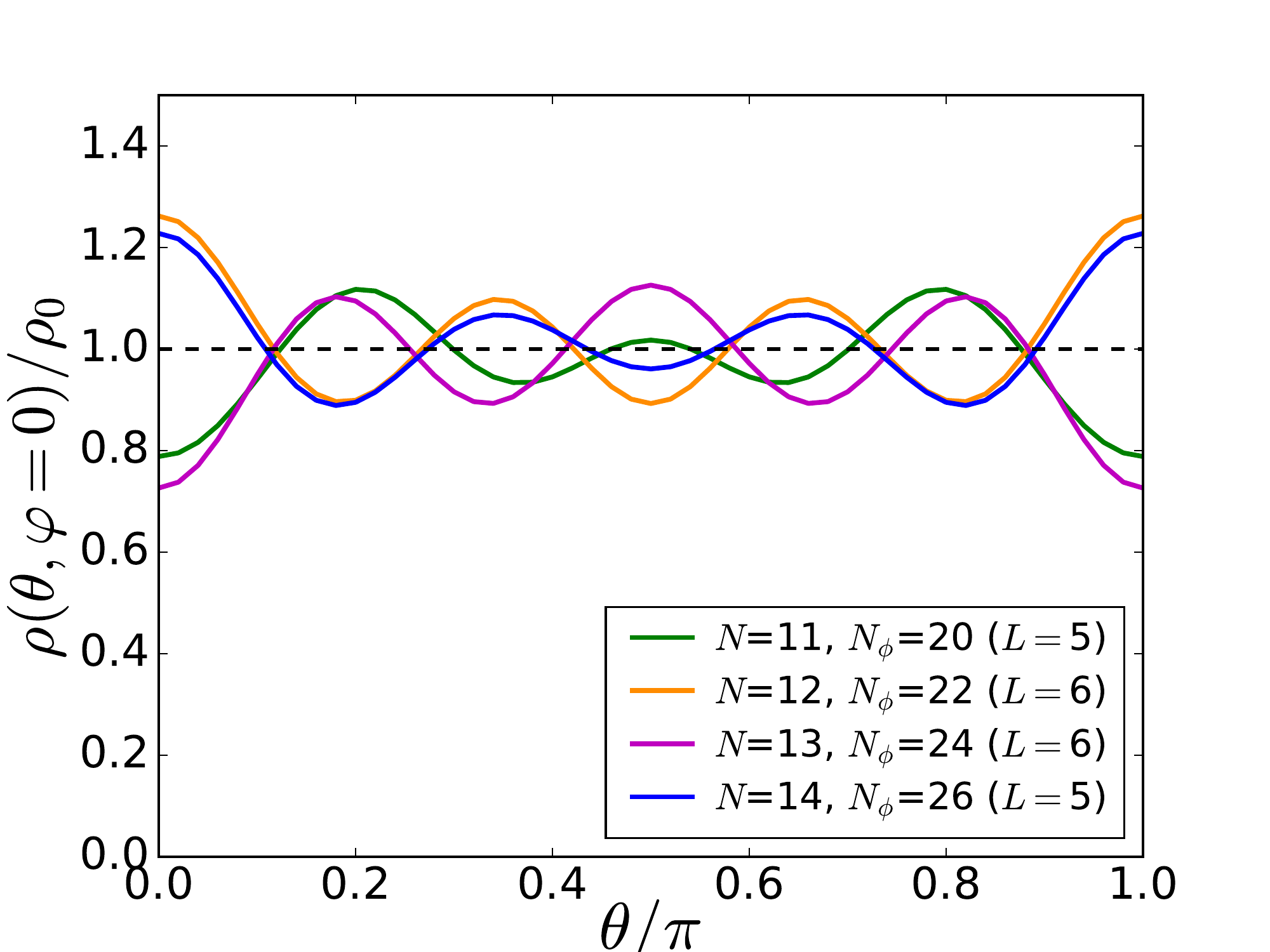}
 \includegraphics[width = 8.5cm]{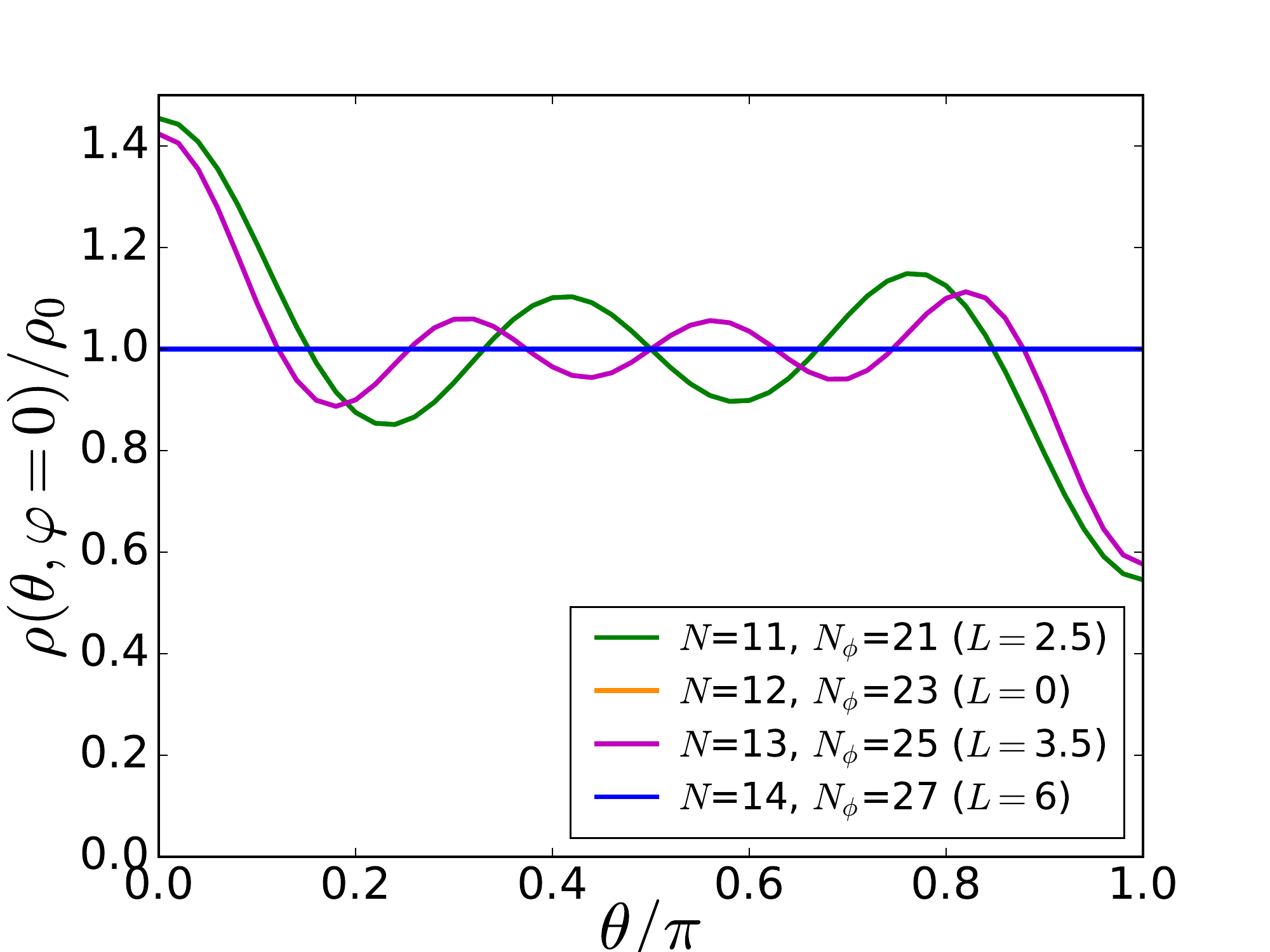}
 \caption{ (color online) 
 Density of fermions in the ground state of the half-filled Landau level in the cases examined in Fig.~\ref{COE_HFLL}. 
 The density $\rho(\theta, \varphi)$ is normalized by the average density $\rho_0 = N/S$, where $S$ is the area of the sphere. 
 The dashed line in the left panel indicates the case of homogeneous density as a guide for eyes. 
 In the right panel, the data for $N=12$ and $14$ are overlapping. 
}
 \label{fig_particle_density}
\end{figure*}

% [ Numerical result ]--------------------
Figures~\ref{Comparison_fig} and~\ref{COE_HFLL} present
the cumulative distribution function $n(\lam)$ and the COE $C_E(T_E)$ (inset of Fig.~\ref{Comparison_fig}) obtained in this way. 
In Fig.~\ref{Comparison_fig}, we compare the numerical data of $n(\lam)$ for $N=14$
with the analytic formulae with $\alpha=2/3$ and $\alpha=1$.
The numerical data clearly show a better agreement with the $\alpha=2/3$ formula than with the $\alpha=1$ one,
in both the Fermi sea case ($N_\phi = 2N-2$) and the particle-hole symmetric case ($N_\phi = 2N-1$).
The data of $C_E(T_E)$ plotted in logarithmic scales again show a variation of slope 
although a rough agreement with $\alpha=2/3$ is found for $10^{-1}\lesssim T_E \lesssim 10^{0}$. 
We furthermore compare the results for different $N$ in Fig.~\ref{COE_HFLL}.
One can see that the data of $n(\lam)$ tend to approach the analytic formula with $\alpha=2/3$ with increasing $N$ 
although a marked deviation from the formula is found for $(N, N_\phi) =(12, 22)$.
We infer that this deviation originates primarily from the spatial inhomogeneity of the fermion density around the boundary of the subregion $A$. 
When the ground state has a nonzero magnitude of the angular momentum $L>0$, 
the $L_z=0$ state used in our calculations can exhibit an inhomogeneous density 
that depends on the azimuthal angle $\theta$ (but not on the polar angle $\varphi$ because of the axial symmetry), as displayed in Fig.~\ref{fig_particle_density}. 
The ground states for $(N, N_\phi) =(12, 22)$ and $(13,24)$ in the Fermi sea case have comparatively large $L$, 
and, as seen in the figure, show appreciable deviations of the density from the average value at the boundary of $A$ ($\theta=\pi/2$, the equator of the sphere). 
Therefore, these states can exhibit large finite-size effects; 
owing to their non-universal nature, such effects are prominent only for $(N, N_\phi) =(12, 22)$ in Fig.~\ref{COE_HFLL}. 
In the particle-hole symmetric case (Fig.~\ref{fig_particle_density}, right), because of a unique antisymmetric behavior, 
the deviation of the density from the average value vanishes at the boundary of $A$, 
which may explain smaller finite-size effects than the Fermi sea case as seen in Fig.~\ref{COE_HFLL}. 

%############################
%############################

%############################
%############################

%% ---------------- Section 4 ---------------------
%%%%%%%%%%%%%%%%%%%%%%%%%%%%%%%%%%%%%%%%%%%%%%%%%
\section{Conclusions\label{Section_Conclusion}}
%%%%%%%%%%%%%%%%%%%%%%%%%%%%%%%%%%%%%%%%%%%%%%%%%

% [ Summary ]--------------------
In this paper, we have studied the COE $C_E(T_E)$ and the cumulative distribution function $n(\lam)$ of RDM eigenvalues in gapless systems. 
Assuming a power-law behavior $C_E\sim T_E^\alpha$ at low $T_E$, we have derived an analytic formula of $n(\lam)$ as in Eq.~\eqref{Formula}. 
We have numerically tested the effectiveness of the formula in relativistic free scalar bosons in two spatial dimensions,  
and find that the distribution of RDM eigenvalues can detect the expected $\alpha=3$ scaling of the COE 
much more efficiently than the raw data of the COE. 
We have also calculated the distribution of RDM eigenvalues in the ground state of the half-filled Landau level with short-range interactions, 
and find a better agreement with the $\alpha=2/3$ formula than with the $\alpha=1$ one, 
which indicates a non-Fermi-liquid nature of the system. 
We have also found that our data tend to approach the $\alpha=2/3$ formula with increasing $N$.
This suggests an intriguing possibility that the COE and the physical heat capacity show the same power-law behavior in this strongly interacting metallic state. 

% [ Discussions ]--------------------
The correspondence between the ES and the physical energy spectrum has been known in gapped topological phases
\cite{KitaevPreskill2006, LiHaldane2008, Thomale2010, Hong2010, TurnerZhangVishwanath2010, Fidkowski2010, Sterdyniak2012,
Dubail2012RSES,QiKatsuraLudwig2012,Chandran2011,Dubail2012,SwingleSenthil2012,Lundgren2013,CanoHughes2015}
and in some gapless systems
\cite{Lauchli2013,MetlitskiGrover2011,Alba2013, Kolley2013}. 
Our numerical result on the half-filled Landau level suggests that this correspondence also holds in a strongly interacting metallic state. 
It would be interesting to investigate whether this correspondence holds in other gapless systems with strong interactions. 
Calculation of the distribution of RDM eigenvalues would be useful for this purpose 
as the comparison with the analytical formula \eqref{Formula} allows us to efficiently probe the low-energy properties of the ES 
as demonstrated in this paper. 
This contrasts with the strategies of Refs.\ \cite{Lauchli2013,MetlitskiGrover2011,Alba2013, Kolley2013}, 
where the ES was compared with the known tower structure of the bulk energy spectrum; 
the use of the distribution of RDM eigenvalues does not require such prior knowledge. 
While the general condition for the correspondence between the ES and the physical spectrum is not known, 
the deviation of the COE or the distribution function from the $\alpha=1$ behavior can already signal a non-Fermi-liquid nature, as explained in Sec.~\ref{Section_Intro}. 
A particularly interesting class of systems to apply this idea are critical spin liquids with a spinon Fermi surface as studied in \cite{ZhangGrover2011,Grover2013}, 
which are also considered to show a heat capacity scaling as $T^{2/3}$ similarly to the half-filled Landau level. 

%% ------------ acknowledgement -----------
\begin{acknowledgments}
  YON thanks Y. Nakaguchi and T. Numasawa for valuable discussions.
  The authors would like to thank hospitality of Yukawa Insitute for Theoretical Physics at Kyoto university
  during the long-term workshop YITP-T-16-01
  ``Quantum Information in String Theory and Many-body Systems'' (May, 2016), where part of our work was done.
  YON was supported by Advanced Leading Graduate Course for Photon Science (ALPS) of
  Japan Society for the Promotion of Science (JSPS) and by JSPS KAKENHI Grant No.\ JP16J01135.
  SF was supported by JSPS KAKENHI Grant No.\ JP25800225 and the Matsuo Foundation. 
\end{acknowledgments}

%% --------------- appendix -----------------------------
\appendix

%%%%%%%%%%%%%%%%%%%%%%%%%%%%%%%%%%%%%%%%%%%%%%%%%
\section{$n_\alpha(\lam)$ as a sum of hypergeometric function \label{Appendix_hypergeo}}
%%%%%%%%%%%%%%%%%%%%%%%%%%%%%%%%%%%%%%%%%%%%%%%%%
When the exponent $\alpha$ of the COE is a rational number,  
the formula $n_\alpha(\lam)$ for the cumulative distribution function in Eq.~\eqref{Formula} 
can be written as a sum of the hypergeometric functions.
In this appendix, we present explicit forms of such expressions.

The hypergeometric function is defined as
\begin{equation}
_p F_q ( \{ a_1, \ldots, a_p \}; \{ b_1, \ldots, b_q \} ; x)
= \sum_{k=0}^\infty \frac{ (a_1)_k \cdots (a_p)_k }{ (b_1)_k \cdots (b_q)_k }  \frac{x^k}{k!},
\end{equation}
where rising factorial $(a_i)_k$ is defined as
$(a_i)_k := a_i (a_i + 1) \cdots (a_i + k -1) = \Gamma(a_i+k) / \Gamma(a_i) $.
When $\alpha$ equals an integer $m \in \mathbb{N}^+= \{1,2, \cdots \}$, one can show that Eq.~\eqref{Formula} reduces to
\begin{align}
 n_m(\lam) = _0 \!\! F_m \lp \{\} ; \left\{ \frac{1}{m}, \ldots, \frac{m-1}{m}, 1 \right\};  \, \frac{x}{m^m} \rp,
\end{align}
where  $x =b  \lp\ln(\lammax/\lam) \rp^m$. %and $b = - \ln \lammax$.
When $\alpha$ is not an integer but a rational number,
$\alpha = p/q$ ($p$ and $q$ are coprime integers),
we observe that Eq.~\eqref{Formula} can be written as the sum of the hypergeometric functions
$_0 F_{M}( g x^q )$ with $x=b  \lp\ln(\lammax/\lam) \rp^\alpha$, 
where a rational number $g$ and an integer $M$ are determined from $p$ and $q$. 
For example, when $\alpha =2/3$ we obtain
\begin{align*}
 n_{2/3}(\lam) = 
 & ~_0F_4\left( \{\} ; \left\{ \frac{1}{3},\frac{1}{2},\frac{2}{3},1 \right\} ; \frac{x^3}{108}\right) \\ + 
  &\frac{x}{\Gamma \left(\frac{5}{3}\right)} ~_0F_4\left( \{\}; \left\{ \frac{2}{3},\frac{5}{6},\frac{4}{3},\frac{4}{3} \right\} ; \frac{x^3}{108}\right) \\ + 
  &\frac{3 x^2}{8 \Gamma \left(\frac{4}{3}\right)} ~_0F_4\left( \{\}; \left\{ \frac{7}{6},\frac{4}{3},\frac{5}{3},\frac{5}{3} \right\} ; \frac{x^3}{108}\right),
\end{align*}
where $x=b  \lp\ln(\lammax/\lam) \rp^{2/3}$.

%%%%%%%%%%%%%%%%%%%%%%%%%%%%%%%%%%%%%%%%%%%%%%%%%
\section{Technical details on numerical calculations in relativistic free scalar bosons %in two spatial dimensions
\label{Appendix_boson}}
%%%%%%%%%%%%%%%%%%%%%%%%%%%%%%%%%%%%%%%%%%%%%%%%%
In Sec.~\ref{subsec_boson},
we calculated the eigenvalues of a RDM for the ground state of relativistic free scaler bosons 
in two spatial dimensions. %with taking as a subregion $A$ a circle centering at the origin. 
This calculation was based on the method of Ref.~\cite{Nakaguchi2015}, 
and we here describe some technical details for completeness.

The field-theory action in continuum is given by  
$S = \int d^2xdt [ (\partial_t \phi)^2 - (\nabla \phi)^2] $. 
We decompose this action in terms of the angular momentum $n$, 
and then discretize the radial direction into $N$ points labeled by $i\in\{1,\ldots,N\}$. 
Setting the lattice constant to unity, 
the resulting Hamiltonian is given by
\begin{equation*}
 H = \frac{1}{2} \sum_{n=-\infty}^\infty \lp  \sum_{i=1}^N \pi_{n,i}^2 + \sum_{i,j=1}^N \phi_{n,i} K^{i,j}_n \phi_{n.j} \rp,
\end{equation*}
where $\phi_{n,i}$ and $\pi_{n,i}$ are the discretized scaler field
and its conjugate field, respectively, for the angular momentum $n$ and the ``site'' $i$. 
The coefficients $K_n^{i,j}$ are given by $ K_n^{1,1} = 3/2 + n^2$, $K_n^{i,i} = 2 + n^2/i^2 \:\: (i \geq 2)$, $K_n^{i,i+1}=K_n^{i+1,i} = -(i+1/2) / \sqrt{i(i+1)}$, and $K_n^{i,j}=0$ (otherwise).

We take as a subregion $A$ a circle of radius $R$ centering at the origin ($0 < R < N$). 
Since the theory is free (quadratic), the RDM of the ground state for the subregion $A$ can be written as a Gibbs state $\rho_A \propto  \exp \lp  - \sum_{k} \epsilon_k b_k^\dagger b_k \rp  $, where
$b_k^\dagger $ and $b_k$ are some bosonic creation and annihilation operators.
The single-particle ``entanglement energy" $\epsilon_k$ can be calculated from eigenvalues
of the correlation matrix $C$ in the subregion $A$~\cite{Peschel2003, Peschel2009, Casini2009}.
The correlation matrix $C$ is defined through the correlation functions in the ground state as $C =\sqrt{XP}$,  where
$X_{i,j} := \otimes_n \langle \phi_{n,i} \phi_{n, j} \rangle_{GS}$ and
$P_{i,j} := \otimes_n \langle \pi_{n, i} \pi_{n, j} \rangle_{GS}$ $(i,j =1, \ldots, R)$.
The eigenvalues of $C$, which we denote by $\xi_k$,  are related with $\epsilon_k$ as
$\frac{1}{2} \mr{coth} \lp \epsilon_k/2 \rp  = \xi_k$.
From $\epsilon_k$, the many-body spectrum of $\rho_A$ and
the distribution function $n(\lam)$ are calculated.

For the data presented in Fig.~\ref{COE_boson2d}, we set $N=40$ and $R=10$.
Since the correlation matrix $C$ is block-diagonalized in terms of the angular momentum $n$, 
$\epsilon_k$'s can be calculated separately for different $n$. 
Since $\epsilon_k$ for large $|n|$ is generally small, we introduce a cutoff $n_\mr{max}$ for $n$ in numerical calculations. 
We checked that the results do not change when we increase $N$ or $n_\mr{max}$ while $R/N$ is fixed.

%%%%%%%%%%%%%%%%%%%%%%%%%%%%%%%%%%%%%%%%%%%%%%%%%
\bibliography{Reference}

\begin{thebibliography}{75}
\bibitem [{not(2017)}]{note_COE_CFT}%
  \BibitemOpen
  \bibinfo {note} {This can be seen by following the logic of
  Sec.~\ref{Section_Derivation} inversely from
  Eq.~\eqref{FormZET}.}\BibitemShut {Stop}%
\end{thebibliography}
%%%%%%%%%%%%%%%%%%%%%%%%%%%%%%%%%%%%%%%%%%%%%%%%%

\end{document}